\pdfoutput=1
\documentclass[aps, reprint, twocolumn,longbibliography,nofootinbib,floatfix]{revtex4-1}
\usepackage[latin1]{inputenc}
\usepackage{graphicx}
\usepackage{amsmath,amssymb,amstext,graphicx,bm,booktabs,hyperref}
\usepackage{fancyhdr}
\usepackage{microtype}

\usepackage{xcolor}
\usepackage{multirow}
\usepackage{placeins}

\newcommand*{\affaddr}[1]{#1}
\newcommand*{\affmark}[1][*]{\textsuperscript{#1}}
\fancypagestyle{alim}{\fancyhf{}\fancyfoot[R]{$^\dagger$Both authors contributed equally and are displayed in random order.}}

\begin{document}

\title{Machine learning enables completely automatic tuning of a quantum device \newline faster than human experts}

\author{H. Moon\affmark[$\dagger$1], D.T. Lennon\affmark[$\dagger$1], J. Kirkpatrick\affmark[2], N.M. van Esbroeck\affmark[1,3], L.C. Camenzind\affmark[4], Liuqi Yu\affmark[4], F. Vigneau\affmark[1], D.M. Zumb\"uhl\affmark[4], G.A.D. Briggs\affmark[1], M.A. Osborne\affmark[5], D. Sejdinovic\affmark[6], E.A. Laird\affmark[7], and N. Ares\affmark[1]\\
\affaddr{\affmark[1]Department of Materials, University of Oxford, Parks Road, Oxford OX1 3PH, United Kingdom}\\
\affaddr{\affmark[2]DeepMind, London EC4 5TW, United Kingdom}\\
\affaddr{\affmark[3]Department of Applied Physics, Eindhoven University of Technology, 5600 MB Eindhoven, The Netherlands}\\
\affaddr{\affmark[4]Department of Physics, University of Basel, 4056 Basel, Switzerland}\\
\affaddr{\affmark[5]Department of Engineering, University of Oxford, Walton Well Road, Oxford OX2 6ED, United Kingdom}\\
\affaddr{\affmark[6]Department of Statistics, University of Oxford
24-29 St Giles, Oxford OX1 3LB, United Kingdom}\\
\affaddr{\affmark[7]Department of Physics, Lancaster University, Lancaster, LA1 4YB, United Kingdom}}

\begin{abstract}
Device variability is a bottleneck for the scalability of semiconductor quantum devices. Increasing device control comes at the cost of a large parameter space that has to be explored in order to find the optimal operating conditions. We demonstrate a statistical tuning algorithm that navigates this entire parameter space, using just a few modelling assumptions, in the search for specific electron transport features. We focused on gate-defined quantum dot devices, demonstrating fully automated tuning of two different devices to double quantum dot regimes in an up to eight-dimensional gate voltage space. We considered a parameter space defined by the maximum range of each gate voltage in these devices, demonstrating expected tuning in under 70 minutes. This performance exceeded a human benchmark, although we recognise that there is room for improvement in the performance of both humans and machines. Our approach is approximately 180 times faster than a pure random search of the parameter space, and it is readily applicable to different material systems and device architectures. With an efficient navigation of the gate voltage space we are able to give a quantitative measurement of device variability, from one device to another and after a thermal cycle of a device. This is a key demonstration of the use of machine learning techniques to explore and optimise the parameter space of quantum devices and overcome the challenge of device variability.

\end{abstract}

\date{\today{}}
\maketitle
\thispagestyle{alim}

Gate defined quantum dots are promising candidates for scalable quantum computation and simulation~\cite{Vandersypen2016,hensgens2017quantum}. They can be completely controlled electrically and are more compact than superconducting qubit implementations~\cite{Vandersypen2016}. These devices operate as transistors, in which electrons are controlled by applied gate voltages. If these gate voltages are set correctly, quantum dots are created, enabling single-electron control. If two such quantum dots are created in close proximity, the double quantum dot can be used to define robust spin qubits from the singlet and triplet states of two electrons~\cite{petta2005coherent,malinowski2017notch}. Due to device variability, caused by charge traps and other device defects, the combination of gate voltage settings which defines a double quantum dot varies unpredictably from device to device, and even in the same device after a thermal cycle. This variability is one of the key challenges that must be overcome in order to create scalable quantum circuits for technological applications such as quantum computing. Typical devices require several gate electrodes, creating a high dimensional parameter space difficult for humans to navigate. Tuning is thus a time consuming activity and we are reaching the limits of our ability to do this manually in arrays of quantum devices. To find, in a multi-dimensional space, the gate voltages which render the device operational is referred to in the literature as coarse tuning~\cite{Teske2019,Botzem2018}.

Here, we present a statistical algorithm which is able to explore the entire multi-dimensional gate voltage space available for electrostatically defined double quantum dots, with the aim of automatically tuning them and studying their variability. Until this work, coarse tuning required manual input~\cite{Baart2016} or was restricted to a small gate voltage subspace~\cite{Kalantre2017}. We demonstrate a completely automated algorithm which is able to tune different devices with up to eight gate electrodes. This is a challenging endeavour because the desired transport features are only present in small regions of the gate-voltage space, surrounded by large regions in which the device is either completely pinched-off or tunnel barriers are too open to observe single electron transport. Moreover, the transport features that indicate the device is tuned to the double dot regime are time-consuming to measure and difficult to parametrise. Machine learning techniques and other automated approaches have been used for tuning quantum devices~\cite{Baart2016,Teske2019,Kalantre2017,Volk2019,VanDiepen2018,Botzem2018,durrer2019automated,mills2019computer,zwolak2019auto}. These techniques are limited to small regions of the device parameter space or require information about the device characteristics. We believe our work significantly improves the state-of-the-art: our algorithm models the entire parameter space and tunes a device completely automatically (without human input), in approximately 70 minutes, faster than the typical tuning by a human expert~\cite{humantuning}. 



\begin{figure}[t]
\centering
\includegraphics[width=\linewidth]{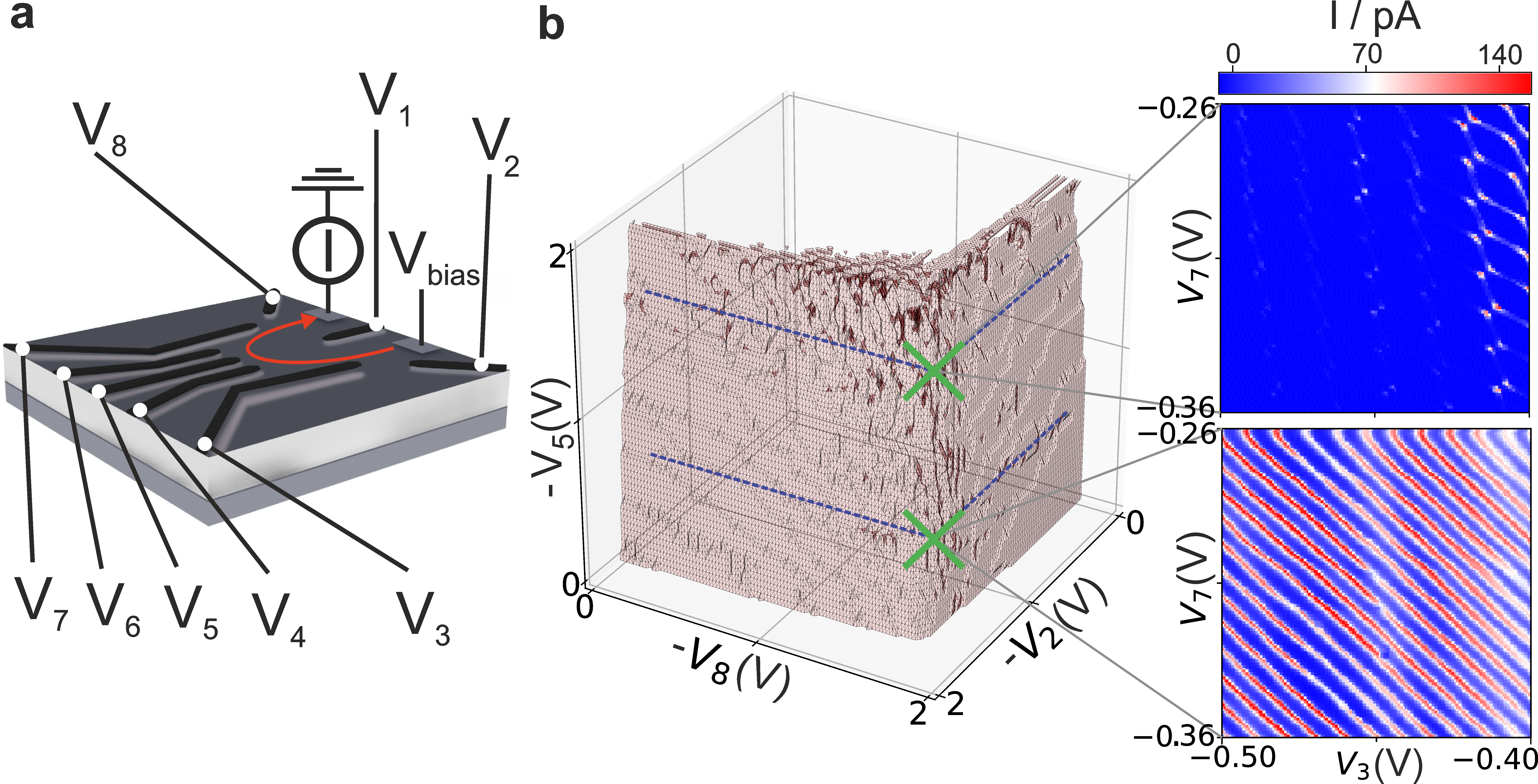}
\caption{\label{fig1}
\textbf{Overview of the device, and gate voltage space.} \textbf{a} Schematic of a gate-defined double quantum dot device. \textbf{b} Left: Boundary hypersurface measured as a function of $V_2$, $V_5$, and $V_8$, with fixed values of $V_1$, $V_3$, $V_4$, $V_6$ and $V_7$.  The current threshold considered to define this hypersurface is 20\% of the maximum measured current. The gate voltage parameter space, restricted to 3D for illustration, contains small regions in which double and single quantum dot transport features can be found. These regions typically appear darker in this representation because they produce complex boundaries. Right: For particular gate voltage coordinates marked with green crosses, the current as a function of $V_7$ and $V_3$ is displayed. The top and bottom current maps display double and single quantum dot transport features, respectively.}
\end{figure}

Our algorithm explores the gate-voltage space by measuring the current flowing through the device, and its design makes only a few assumptions, allowing it to be readily applied to other device architectures. Our quantum dot devices are defined in a 2-dimensional electron gas in a GaAs/AlGaAs heterostructure by Ti/Au gate electrodes. DC voltages applied to these gate electrodes, $V_1$ to $V_8$, create a lateral confinement potential for electrons. A bias voltage $V_{\text{bias}}$ is applied to ohmic contacts to drive a current ($I$) through the device. The device schematic, designed for precise control of the confinement potential~\cite{camenzind2019spectroscopy,stano2019orbital,camenzind2018hyperfine}, is shown in Fig.~\ref{fig1}a. Measurements were performed at $50$~mK.

We consider the space defined by up to eight gate voltages between 0 and -2~V. This range was chosen to avoid leakage currents. In this parameter space, the algorithm has to find the desirable transport features within tens of mV. Identifying these features is slow, as it requires to measure $I$ as a function of two plunger gate voltages, i.e. gate voltages that significantly modify the electron occupation of each quantum dot. Note that fast readout techniques are only suited for small regions of the parameter space. Our algorithm is thus designed to minimize the number of two dimensional current maps required. We make two observations. Firstly, that for very negative gate voltages, no current will flow through the device, i.e. the device is pinched-off. Conversely, for very positive gate voltages, full current will flow and single electron transport will not be achieved. This means that transport features are expected to be found on the hypersurface that separates low and high current regions in parameter space. The second observation is that to achieve single-electron transport, a confinement potential is needed. The particular transport features that evidence single-electron transport are Coulomb peaks, which are peaks in the current flowing through the device as a function of a single plunger gate voltage. These observations leads us to only two modelling assumptions: (i) single and double quantum dot transport features are embedded in a boundary hypersurface, shown in Fig.~\ref{fig1}b, which separates regions in which a measurable current flows, from regions in which the current vanishes; (ii) large regions of this hypersurface do not display transport features.

\begin{figure}[t]
\centering
\includegraphics[width=\linewidth]{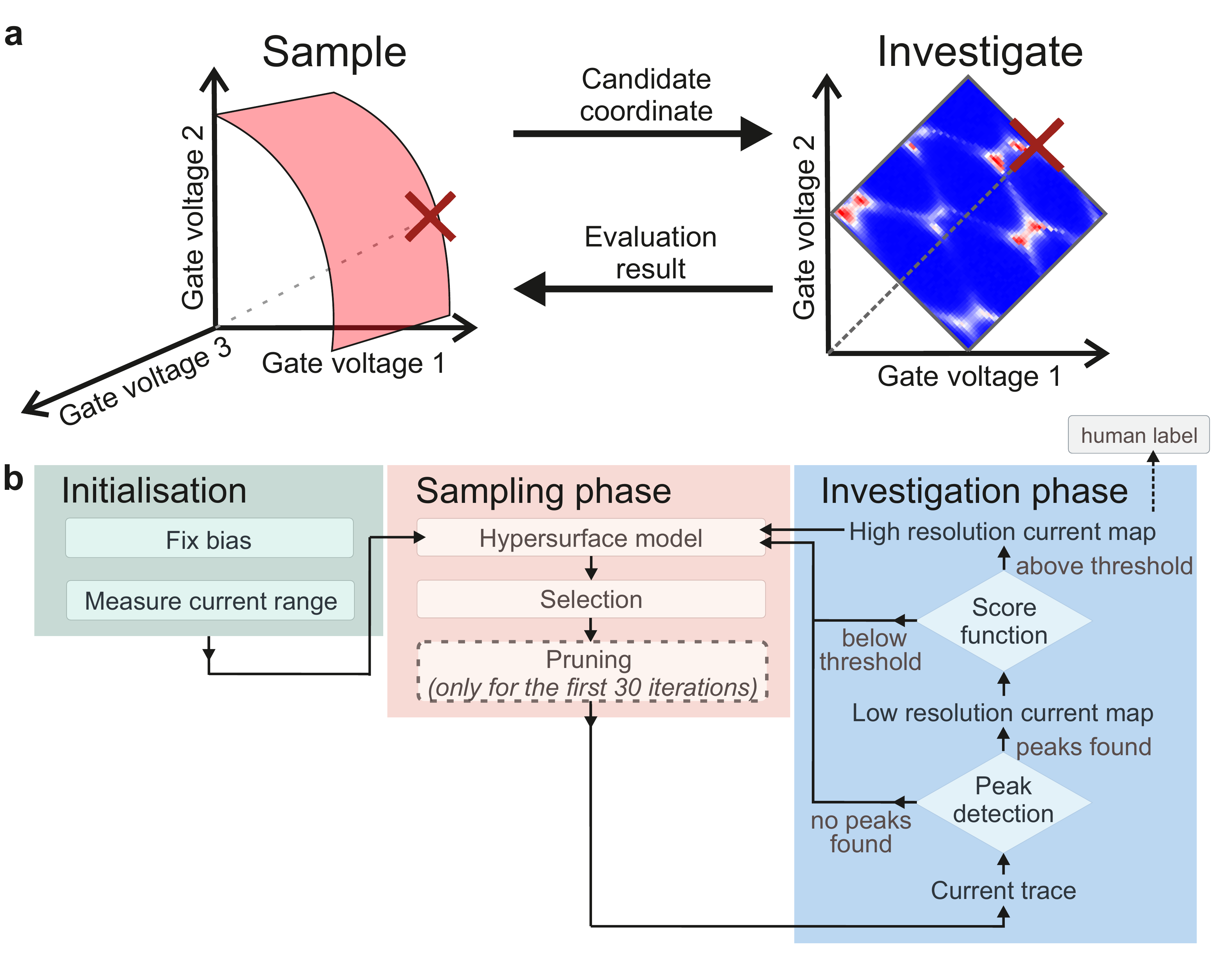}
\caption{\label{fig1b}
\textbf{Overview of the algorithm.} \textbf{a} The sampling phase produces candidate coordinates in gate voltage space, which are on the boundary hypersurface (pink surface). The distance between a candidate coordinate (red cross) and the origin of the gate voltage space is marked with a dashed line. The investigation phase, evaluates the local region by, for example, measuring current maps which are evaluated by a score function. Evaluation results are fed back to the sampling phase.~\textbf{b} Algorithm's block diagram. The algorithm is initialized by fixing $V_{\text{bias}}$ and performing a current measurement at the two extremes of the gate voltage space. The next phase is sampling, in which a candidate coordinate is first produced from a modelled hypersuface, selected and then found on the true hypersurface by performing a current measurement. Also, pruning of unusable regions of the gate voltage space is performed. The investigation phase first detects Coulomb peaks in the current measurement, and if successful, then the Coulomb peak spacing defines a gate voltage window to produce a current map at low resolution. The algorithm increases the scan resolution if the low resolution current map is promising in terms of double quantum dot transport features according to the score function. The result of the investigation phase is fed to the sampling phase after the high resolution scan, or earlier if no Coulomb peaks are detected or if the low resolution scan does not pass the threshold score. Humans label the high resolution map to confirm the presence of features corresponding to the double quantum dot regime.}
\end{figure}

The algorithm consists of two parts: a sampler that generates candidate coordinates on the hypersurface, and an investigation phase in which we collect data in the vicinity of each candidate coordinate to evaluate transport features (Fig.~\ref{fig1b}a). The result of the investigation phase feeds back into the sampler, which chooses a new candidate coordinate in the light of this information. The objective of the sampler is to produce candidate coordinates in gate voltage space for which the device operates as a double quantum dot. A block diagram of the algorithm is displayed in Fig.~\ref{fig1b}b. Our modelling assumptions are based on the physics of gate defined devices leading to minimal constraints; we do not assume a particular shape for the hypersurface, and we instead allow measurements to define it by fitting the data with a Gaussian process. 

We demonstrate over several runs, in two different devices and over thermal cycles, that the algorithm is able to find transport features corresponding to double quantum dots. We perform an ablation study, which identifies the relative contribution of each of the modules that constitute the algorithm, justifying its design. Finally, we demonstrate that our algorithm is capable of quantifying device variability, which has only been theoretically explored so far~\cite{stopa1996quantum}.

Automating experimental science has the impact to significantly accelerate the process of scientific discovery. In this work we show that a combination of simple physical principles and flexible probabilistic machine learning models can be used to efficiently characterise and tune a device. We envisage that in the near future such judicious application of machine learning will have tremendous impact even in areas where only small amounts of data are available and no clear fitness functions can be defined.

\section{Sampling phase}
\label{section:sampling}
The algorithm starts by fixing $V_{\text{bias}}$ and performing a measurement of current at the two extremes of the gate voltage space, $V_i=0$ and $V_i=-2~$V for $i=0,...,N$ where $N$ indicates the number of gate electrodes. This initial calibration sets the current scale that the algorithm requires to define the hypersurface. The search range was chosen to span the typical gate voltage settings for tuning lateral quantum dots, which are usually a few hundred mV. Once this initialization phase is completed, the sampling phase begins. The aim of the sampler is to propose candidate gate voltage coordinates to be evaluated by the investigation phase described in section~\ref{section:Invest}. The sampler is very general, just relying on our two modelling assumptions, as it models the hyper-surface (i), and prunes regions of the hyper-surface that do not contain transport features (ii). For pruning, feedback from the investigation phase is required to evaluate the presence of transport features in a given region of the hypersurface. The sampler learns from this feedback where desired transport features are likely to be found on the hypersurface.

\subsection{Hypersurface model}
\label{section:hypersurface}
\begin{figure*}
\centering
\includegraphics{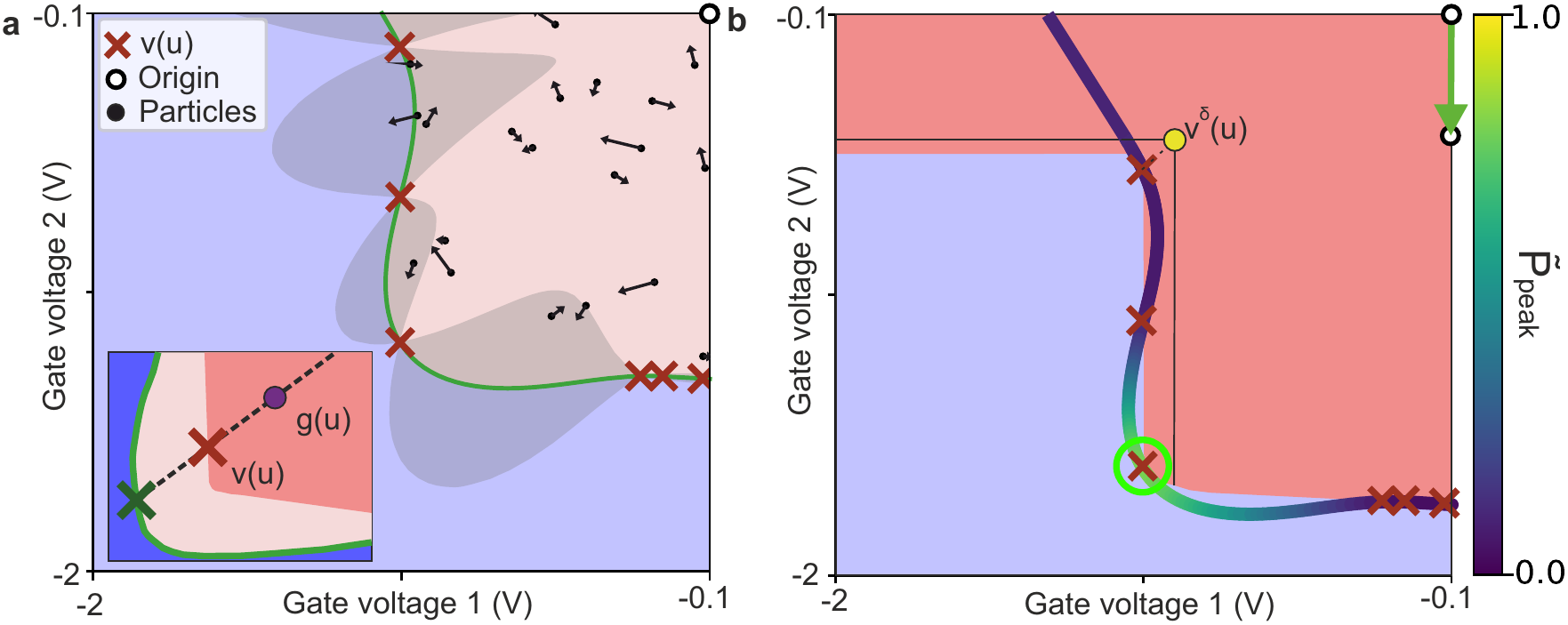}
\caption{\label{fig2}
\textbf{Sampling phase.} 
The gate voltage space is restricted to two dimensions for illustration, and its origin is marked with a circle. The origin does not coincide with 0 gate voltages so that its surroundings are included in the considered gate voltage range. \textbf{a} Schematic hypersurface model. 
The predictions of the model are schematically indicated. Blue (pink) areas represent regions of predicted zero (non zero) current, and the green line represents the predicted hypersurface with uncertainty bounds. These uncertainty bounds vanish for each measured coordinate on the hypersurface. The arrows represent Brownian particles simulated to achieve uniform sampling of the hypersurface.
Inset: Dark pink indicates the true non-zero current region and lighter pink the same region predicted by the model. The algorithm verifies the model prediction (green cross) by measuring the current starting from a coordinate $g(\mathbf{u})$, chosen with high probability in a region of high current, and stops when the true hypersurface is encountered (red cross).
\textbf{b} Selection and pruning. Dark pink indicates a schematic non-zero current region in a blue background representing currents close to zero. The dashed line shows the sweep in gate voltages from a coordinate on the hypersurface to a higher-current coordinate  $\mathbf{v}^\delta(\mathbf{u})$. From $\mathbf{v}^\delta(\mathbf{u})$, black lines indicate current traces as a function of decreasing gate voltages 1 and 2. In this example, pinch-off is not observed as a function of gate voltage 1. The algorithm will then move the origin to match the gate voltage 2 component of $\mathbf{v}^\delta(\mathbf{u})$. 
The multi-colored line represents the estimated probability of finding transport features in the vicinity of a coordinate on the hyper-surface.
}

\end{figure*}


To parametrise the hypersurface, we define the origin $\mathbf{o}$ of the gate voltage space as $V_i=-100$~mV for $i=0,..., N$. This is to allow for measurements around $\mathbf{o}$ while remaining within the defined gate voltage range. We also indicate directions in gate voltage space with a unit vector $\mathbf{u}$. We define the distance from $\mathbf{o}$ to the hypersurface for a given direction $\mathbf{u}$ as $r(\mathbf{u})$. The corresponding coordinate on the hypersurface is $\mathbf{v}(\mathbf{u})=r(\mathbf{u})\mathbf{u}+\mathbf{o}$. From measured distances $\lbrace r(\mathbf{u}_i) \rbrace_{i=1,\ldots, n}$ in $n$ sampling iterations, the algorithm predicts the distance $\tilde{r}(\mathbf{u})$ for every $\mathbf{u}$ using a Gaussian process regression: $\tilde{r}(\mathbf{u})\sim\mathcal{N}(m(\mathbf{u}), s^2(\mathbf{u}))$, i.e. $\tilde{r}(\mathbf{u})$ follows a Gaussian distribution $\mathcal{N}$ with $m(\mathbf{u})$ and $s^2(\mathbf{u})$ the posterior mean and variance, respectively (Fig.~\ref{fig2}a). We define $\mathbf{\tilde{v}}(\mathbf{u})=\tilde{r}(\mathbf{u})\mathbf{u}+\mathbf{o}$ the hypersurface coordinate predicted by the model in direction $\mathbf{u}$. Details of the prediction models can be found in the Supplementary Material.

For a randomly generated direction $\mathbf{u}$, the algorithm verifies this prediction by measuring the current along $\mathbf{u}$. This current measurement does not commence at $\mathbf{o}$, since it is time-consuming to sweep a big range of gate voltages. It instead starts from a coordinate $\mathbf{g}(\mathbf{u})=(m(\mathbf{u})-2s(\mathbf{u}))\mathbf{u}+\mathbf{o}$, which is with high probability in a region of high current, and stops when the true hypersurface coordinate $\mathbf{v}(\mathbf{u})$ is found. The coordinate $\mathbf{g}(\mathbf{u})$ is checked to be always further away from $V_i=0$~mV than $\mathbf{o}$. If this coordinate were to be in a pinched-off region, the algorithm measures the current along $\mathbf{-u}$ and when high values of current are found, it reverts to measuring along $\mathbf{u}$ to find $\mathbf{v}(\mathbf{u})$. 

To sample the hypersurface uniformly, we cannot rely on uniformly distributed random directions of $\mathbf{u}$, because the resulting set of predictions $\mathbf{\tilde{v}}(\mathbf{u})$ will be denser in regions of the hypersurface for which $\tilde{r}(\mathbf{u})$ is shorter. To achieve approximately uniform sampling, we simulate particles undergoing Brownian motion inside the volume delimited by the modelled hypersurface $m(\mathbf{u})$. The algorithm collects the coordinates on this hypersurface which the particles hit. We have developed this Brownian motion approach based on the findings from \cite{Narayanan2008}, where an algorithm for generating random samples from geometric objects is established. 



\subsection{Selection and pruning}
\label{section:Selection}
From the set $\lbrace \tilde{\mathbf{v}} \rbrace$ collected by the Brownian particle simulation, one coordinate is chosen according to the estimated probability of finding transport features in the coordinate's vicinity. The particular transport features we are interested in are sets of Coulomb peaks evidencing single-electron tunneling. The algorithm uses Gaussian process classification~\cite{Rasmussen2005} to estimate the probability of measuring Coulomb peaks $\tilde{P}_\textrm{peak}(\mathbf{v})$ in the vicinity of a given $\mathbf{v}$. When selecting a coordinate to sample we must make a decision about whether to exploit our prediction of $\tilde{P}_\textrm{peak}(\mathbf{v})$ (by choosing the maximum of $\tilde{P}_\textrm{peak}(\mathbf{v})$) or whether to explore the parameter space in order to better predict $\tilde{P}_\textrm{peak}(\mathbf{v})$. To balance this trade-off we utilise Thompson sampling which selects candidate coordinates from $\lbrace \tilde{\mathbf{v}} \rbrace$ proportionally to their predicted probability $\tilde{P}_\textrm{peak}(\tilde{\mathbf{v}})$.

We have also introduced a pruning rule to rapidly exclude directions $\mathbf{u}$ for which the hypersurface cannot be intersected, since it lies outside the gate voltage range considered. The algorithm takes a pinch-off coordinate $\mathbf{v}(\mathbf{u})$, increases all gate voltages by a step-back parameter, and takes current traces along the gate voltage axes (in direction of decreasing gate voltages) to look for pinch-off. 
Because the step-back vector $\boldsymbol\delta$ has all positive elements, the starting point of these traces $\mathbf{v}^\delta(\mathbf{u})=\mathbf{v}(\mathbf{u}) + \boldsymbol\delta$ will probably be located in a high current region. If a pinch-off coordinate can only be found along one gate voltage axis $k$, then the algorithm moves the $k$ component of the origin $\mathbf{o}$ to match the $k$ component of $\mathbf{v}^\delta(\mathbf{u})$ (Fig.~\ref{fig2}b). This pruning rule excludes very quickly those regions of the parameter space for which the  hypersurface cannot be intersected within the considered gate voltage range, and thus it is just applied for the first 30 iteration of the algorithm. We chose the step-back parameter as 100~mV for all experiments. Because $\mathbf{o}$ was chosen to allow for this step back, $\mathbf{v}^\delta(\mathbf{u})$ to always lie in the considered gate voltage range.

\section{Investigation phase}
\label{section:Invest}

The aim of the investigation phase is to evaluate the local region of a candidate coordinate to determine if desired transport features are present, and provide feedback to the sampling phase about its findings. It is important to obtain a fast but reliable evaluation of candidate coordinates. The algorithm first performs a quick measurement to determine if there are Coulomb peaks in the vicinity of a given candidate coordinate. This peak detection test will inform the selection of candidate coordinates at the sampling phase.

If Coulomb peaks are found in the vicinity of a candidate
coordinate, indicating that it is in a quantum dot regime, the algorithm further explores the region to determine whether it is in a double quantum dot regime, and if so produces a score ranking its quality.




\subsection{Peak detection}
\label{section:peakdetection}
For a given pinch-off coordinate $\mathbf{v}(\mathbf{u})$, the algorithm explores the plane containing this coordinate and defined by directions in gate voltage space $\hat{V}_3$ and $\hat{V}_7$, chosen as increasing plunger gate voltages $V_3$ and $V_7$.
The first measurement made by the investigation phase is a diagonal trace of current in this plane, i.e. in direction $\hat{V}_e = \frac{1}{\sqrt{2}}(\hat{V}_3 + \hat{V}_7)$ opposite to pinch-off, starting from $\mathbf{v}(\mathbf{u})$. The trace is $128$~mV long in gate voltage space and the voltage resolution is 1~mV. The voltage range and resolution are chosen parameters. Peak detection is used to determine the presence of Coulomb peaks in this trace. If the current trace does not reveal peaks, investigation of this coordinate $\mathbf{v}(\mathbf{u})$ ceases, and the sampler generates new candidate coordinate in the light of this information. If peaks are observed then the investigation phase progresses to further exploration of the plane in question. 


\subsection{Score function decision}
\label{section:score}

If a Coulomb peak was observed at the peak detection stage, the algorithm proceeds to explore further the vicinity of the candidate coordinate. The average spacing between observed peaks is used to define a window in the plane given by $\hat{V}_e$ and the orthogonal vector $\hat{V}_a =\frac{1}{\sqrt{2}}(\hat{V}_3 - \hat{V}_7)$. This window in gate voltage space is square and it is bounded in direction $\hat{V}_e$ by the pinch-off coordinate $\mathbf{v}(\mathbf{u})$. The window's size is 3.5 times the average spacing between peaks. If less than three peaks are observed, the window's default size is chosen to be 100~mV to observe other peaks if present. A current map in this window can be measured with different resolutions. We define low (high) resolution as $16\times16$ ($48\times48$) pixels in the window.



With a low resolution current map, the algorithm evaluates the quantum dot regime using a score function. This score function is a predefined mathematical expression designed to reward specific transport features. A honeycomb pattern observed in the current map indicates that the device is tuned to the double quantum dot regime. The score function thus rewards current maps in which sharp and curved lines are observed. A detailed description of the score function can be found in the Supplementary Material.

Based on this score, the algorithm makes a decision. If the score is bigger than or equal to a certain threshold, the current map is measured again with high resolution for a human to check if the device is indeed tuned to the double quantum dot regime. If the score is below the threshold, the algorithm reverts to the sampling phase. In this way, just current maps in which a honeycomb pattern is observed are selected (see appendix~\ref{classifier}). For our experiments we dynamically adjust the threshold during the experiment such that the algorithm proceeds to measure at high resolution only $15\%$ of the low resolution current maps. The statistical analysis on the optimal score threshold can be found in the Supplementary Material.

\section{Results}

The performance of our algorithm is assessed by a statistical analysis of the expected success time $\mu_t$. This is defined as the time it takes the algorithm to acquire a high-resolution current map that is confirmed a posteriori by humans as containing double quantum dot features. Because human labelling is subjective, three different researchers labelled all current maps, deciding in each case if they could identify features corresponding to the double quantum dot regime, with no other information available. See Supplementary Material for details of the multi-labeller statistical analysis.

\begin{figure}[!htbp]
\centering
\includegraphics[width=\linewidth]{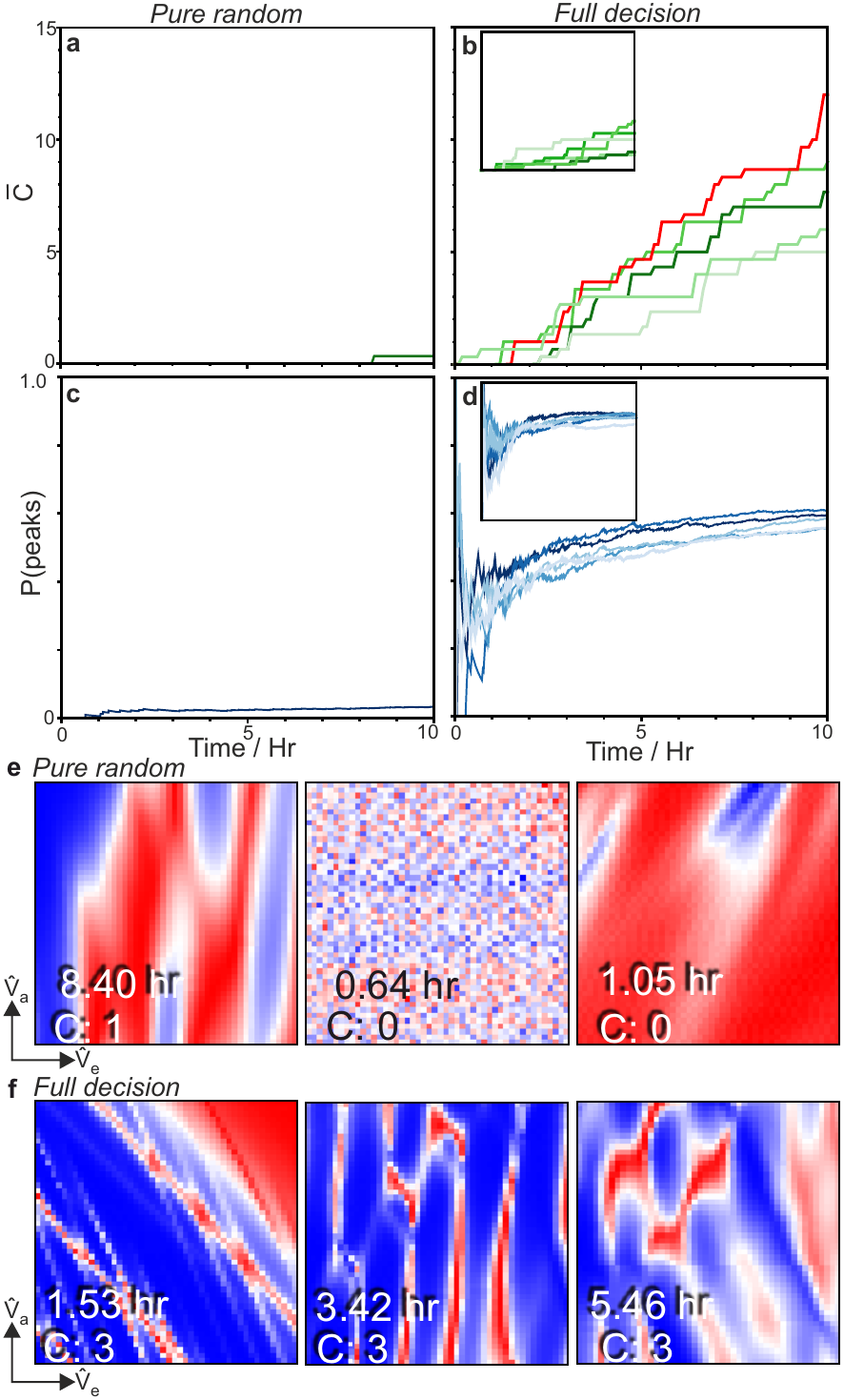}
\caption{\label{fig3} \textbf{Algorithm's performance.} \textbf{a-d} Average number of current maps displaying double quantum dot features, $\bar{C}$, and $P(\textrm{peaks})$ as a function of laboratory time. Current maps are labelled by humans a posteriori, i.e. after the algorithm is stopped. \textbf{a,c} and \textbf{b,d} correspond to one run of \textit{Pure random} and five runs of our algorithm, respectively. All algorithm runs displayed in main panels were performed in Device 2, while insets show runs of our algorithm in Device 1. \textbf{e,f} High resolution current maps produced by \textit{Pure random} and one of our algorithm runs, respectively. We indicate the laboratory time at which they were acquired and the number of labellers, $C$, that identified them as displaying double quantum dot features. Current maps are ordered from left to right in decreasing order of $C$, and maps that have identical values of $C$ are displayed in the order at which they were sampled. Each panel uses an independent colour scale running from red (highest current measured) to blue (lowest current).}
\end{figure}

\subsection{Device tuning}
\label{subsec:result_tuning}

To benchmark the tuning speed of our algorithm, we ran it several times on two different devices with identical gate architecture, Devices 1 and 2, and we compared its performance with a \textit{Pure random} algorithm. The \textit{Pure random} algorithm searches the whole gate voltage parameter space by producing a uniform distribution of candidate coordinates. It does not include a hypersurface model or pruning rules, but uses peak detection in its investigation phase. 

As mentioned in the introduction, we consider a gate voltage space whose dimension is defined by the number of working gate electrodes, and we provide a gate voltage range that avoids leakage currents. While for Device 1 we considered the eight-dimensional parameter space defined by all its gate electrodes, for Device 2 we excluded gate electrode 6 by setting $V_6=0$~mV due to observed leakage currents associated with this gate. 

We define the average count $\bar{C}$ as the number of current maps labelled by humans as displaying double quantum dot features divided by the number of labellers. For a run of the \textit{Pure random} algorithm in Device 2 and five runs of our algorithm in Devices 1 and 2, we calculated $\bar{C}$ as a function of laboratory time (Fig.~\ref{fig3}a,b). We observe that $\bar{C}$ is vastly superior for our algorithm compared with \textit{Pure random}, illustrating the magnitude of the parameter space.

The labellers considered a total of 2048 current maps produced in different runs, including those of the ablation study in Section ~\ref{subsec:result_abl}. The labellers had no information of the run in which each current map was produced or the algorithm used. For the \textit{Pure random} approach, the labelled set was composed of 51 current maps produced by the algorithm and 100 randomly selected from the set of 2048.

The time $\mu_t$ is estimated by the multi-labeller statistics. The multi-labeller statistics uses an average likelihood of $\mu_t$ over multiple labellers and produces an aggregated posterior distribution (see Supplementary Material). From this distribution, the median and 80\% (equal-tailed) credible interval of $\mu_t$ is 2.8 hr and (1.9, 7.3) hr for Device 1 and 1.1 hr and (0.9, 1.6) hr for Device 2. Experienced humans require approximately 3 hr to tune a device of similar characteristics into exhibiting double quantum dot features~\cite{humantuning}. Our algorithm's performance might therefore be considered super human. Due to device variability, the hypersurfaces of these two devices are significantly different, showing our algorithm is capable of coping with those differences.


In Fig.~\ref{fig3}c,d, we compare the probability of measuring Coulomb peaks in the vicinity of a given $\mathbf{v}(\mathbf{u})$, $P(\textrm{peaks})$, for \textit{Pure random} and different runs of our algorithm. We calculate $P(\textrm{peaks})$ as the number of sampled coordinates in the vicinity of which Coulomb peaks were detected over $n$. In this way, we confirm that $P(\textrm{peaks})$ is significantly increased by our algorithm. It has a rapid growth followed by saturation. 

Figure~\ref{fig3}e,f shows the high resolution current maps produced by \textit{Pure random} and one of our algorithm runs. We observe that our algorithm produces high resolution current maps which are recognized by all labellers as displaying double quantum dot features within 1.53 hr. The three current maps in Figure~\ref{fig3}f correspond to double quantum dot regimes found by our algorithm in different regions of the gate voltage space. The number of labellers $C$ who identify the current maps produced by \textit{Pure random} as corresponding to double quantum dots, $C$, is 0 or 1. 


To significantly reduce tuning times, we then modified our algorithm to group gate electrodes that perform similar functions. The algorithm assigns equal gate voltages to gate electrodes in the same group. 
For Device 1, we organized the eight gate electrodes into four groups: $G_1 = (V_1)$, $G_2 = (V_2,V_8)$, $G_3 = (V_3,V_7)$, and $G_4 = (V_4,V_5,V_6)$. In this case, the median and 80\% credible interval of $\mu_t$ improve to 0.6 hr and (0.4, 1.1) hr. This approach, by exploiting knowledge of the device architecture, reduces $\mu_t$ by more than four times.

\begin{figure*}
\centering
\includegraphics{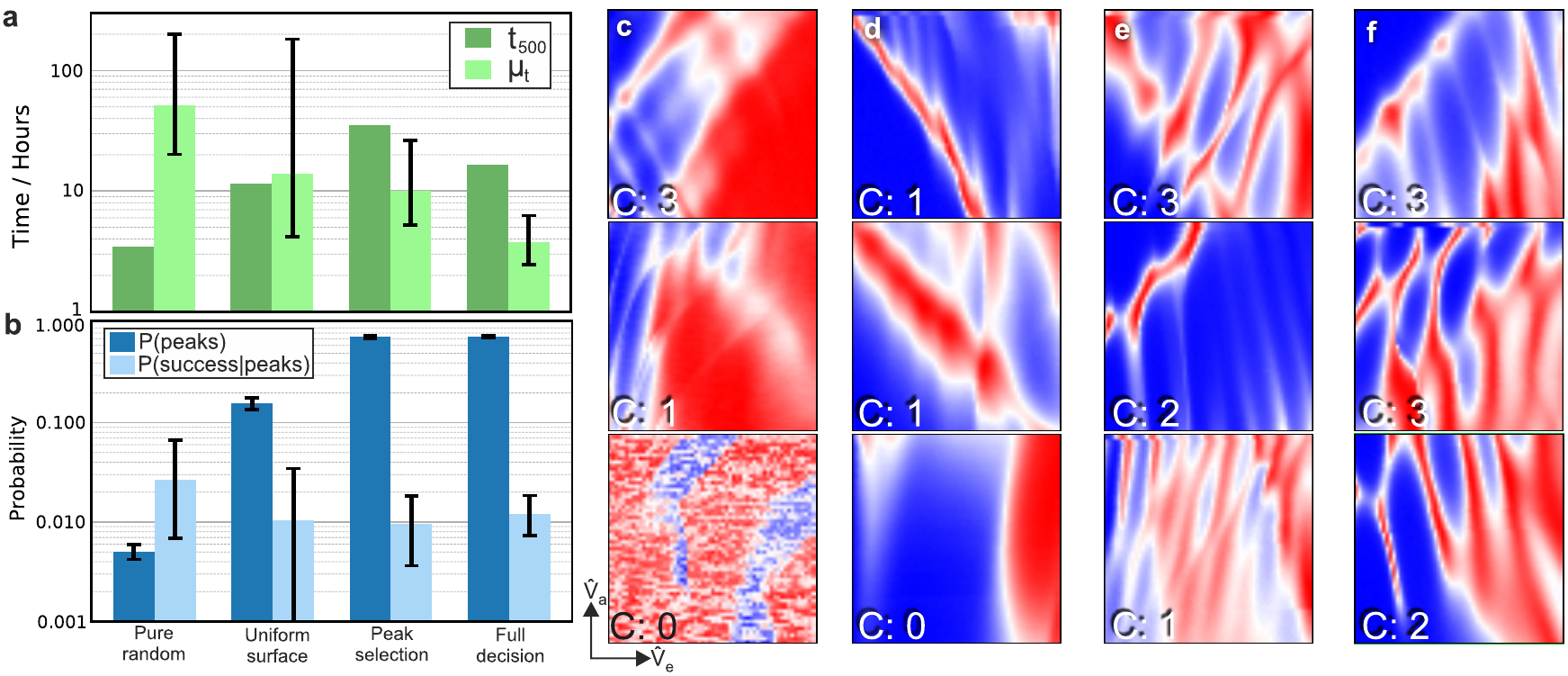}
\caption{\label{fig4}
\textbf{Ablation study.} \textbf{a} and \textbf{b} bar charts comparing $\mu$ (light green), $t_{\text{500}}$ (dark green), $P(\textrm{peaks})$ (dark blue) and $P(\textrm{success}|\textrm{peaks})$ (light blue) for the different algorithms considered. Error bars represent $80\%$ (equal-tailed) credible intervals. Due to a measurement problem, 459 sampling iterations instead of 500 were considered for the \textit{Full decision} algorithm. \textbf{c} to \textbf{f} High resolution current maps sampled by \textit{Pure random}, \textit{Uniform surface}, \textit{Peak selection} and \textit{Full decision}, respectively. In each panel, we indicate $C$, the number of human labellers that identified a map as displaying double quantum dot features. Current maps with identical values of $C$ are displayed in the order in which they were sampled, from top to bottom. Current maps with $C=0$ were randomly selected. Each panel uses an independent colour scale running from red (highest current measured) to blue (lowest current).
}
\end{figure*}
\subsection{Ablation study}
\label{subsec:result_abl}

Our algorithm combines a sampling phase, which integrates the hypersurface model (section \ref{section:hypersurface}) with selection and pruning (section \ref{section:Selection}), and an investigation phase that includes peak detection (section \ref{section:peakdetection}) and score function decisions (section \ref{section:score}). Each of these modules contributes to the algorithm's performance. An ablation study identifies the relative contributions of each module, justifying the algorithm's architecture. For this ablation study we chose to compare our algorithm, \textit{Full decision}, with three reduced versions that combine different modules; \textit{Pure random}, \textit{Uniform surface} and \textit{Peak selection} (see Table \ref{table:table1}). 

\begin{table*}
\centering
\caption{\label{table:table1}Comparison of algorithms used in the ablation study.}
\begin{ruledtabular}
\begin{tabular*}{\textwidth}{@{\extracolsep{\fill}}lcccc}
\multirow{2}{4em}{Algorithm} &  \multicolumn{2}{c}{Sampling phase} & \multicolumn{2}{c}{Investigation phase}\\
\cline{2-3}
\cline{4-5}
{} & hypersurface model& Selection and pruning & Peak detection & Score function decision\\
\hline
Pure random & $\times$ & $\times$ & \checkmark & $\times$ \\
Uniform surface & \checkmark & $\times$ & \checkmark & $\times$ \\
Peak selection & \checkmark & \checkmark & \checkmark & $\times$ \\
Full decision & \checkmark & \checkmark & \checkmark & \checkmark \\
\end{tabular*}
\end{ruledtabular}
\end{table*}

\textit{Pure random}, defined in the previous section, produces a uniform distribution of candidate coordinates over the whole gate voltage space. It excludes the hypersurface model and pruning rules. \textit{Uniform surface} makes use of the hypersurface model, but no selection or pruning rules are considered. \textit{Peak selection} combines the hypersurface model with selection and pruning rules. These three algorithms use peak detection in their investigation phase, but none of them use the score function decision. For the ablation study, we define low (high) resolution as $20\times20$ ($60\times60$) pixels.



To analyse the algorithm's performance, we estimate $P(\textrm{peaks})$ and the probability of success, i.e. the probability to acquire a high-resolution current map labelled as containing double quantum dot features, given Coulomb peak measurements $P(\textrm{success}|\textrm{peaks})$. To take measurement times into consideration, we define $t_{500}$ as the time to sample and investigate 500 coordinates in gate voltage space. The ablation study was performed in Device 1 keeping investigation phase parameters fixed. Results are displayed in Fig.~\ref{fig4}. 

Figure~\ref{fig4}a shows that the introduction of the hypersurface model, and selection and pruning, increases $t_{500}$. This is because $P(\textrm{peaks})$ increases with these modules  (Fig.~\ref{fig4}b), and thus the number of low and high resolution current maps required by the investigation phase is larger. Within uncertainty, $P(\textrm{success}|\textrm{peaks})$ remains mostly constant for the different algorithms considered. The result is a decreasing $\mu_t$ from \textit{Pure random} to \textit{Peak selection} within experimental uncertainty. See appendix~\ref{ablation_appendix} for a mathematical analysis of these results.

The reason behind the use of peak detection in all the algorithms considered for this ablation study is the vast amount of measurement time that would have been required otherwise. Without peak detection, the posterior median estimate of $\mu_t$ for \textit{Pure random} is 680 hr.


To complete the ablation study, we compare the considered algorithms with the grouped gates approach described in the previous section, keeping parameters such as the current map resolutions are equal. We found $\mu_t=80.5$~mins (see Supplementary Material).




In summary, comparing \textit{Pure random} and \textit{Uniform surface}, we show the importance of hypersurface modelling. The difference between \textit{Uniform surface} and \textit{Peak selection} highlights the importance of selection and pruning. The improved performance of \textit{Full decision} with respect to \textit{Peak selection} evidences the tuning speedup achieved by the introduction of the score function. These results demonstrate \textit{Full decision} exhibits the shortest $\mu_t$ and imply an improvement over \textit{Pure random} without peak detection of approximately 180 times.


\section{Device variability}
\label{section:variability}
The variability of electrostatically defined quantum devices has not been quantitatively studied so far. We have been able to exploit our algorithms for this purpose. Using the uniform surface algorithm only (no investigation phase), we obtain a set of coordinates on the hyper-surface $\mathbf{v_a}$. Changes occurring to this hyper-surface are detected by running the algorithm again and comparing the new set of coordinates, $\mathbf{v_b}$, with $\mathbf{v_a}$. This comparison can be done by a point set registration method, which allows us to find a transformation between point sets, i.e. between sets of coordinates. 

Affine transformations have proven adequate to find useful combinations of gate voltages for device tuning~\cite{Volk2019,VanDiepen2018}. To find a measure of device variability, understood as changes occurring to a device's hyper-surface, we thus use an affine transformation $\mathbf{v_t} = B\mathbf{v_b}$, with $B$ a matrix which is a function of the transformation's parameters. We are looking for a transformation of coordinates that converts $\mathbf{v_b}$ into a set of coordinates $\mathbf{v_t}$ which is as similar to $\mathbf{v_a}$ as possible.
 
The particular point set registration method we used is coherent point drift registration ~\cite{Myronenko2010}. This method works with an affine transformation which includes a translation vector. We have modified the method to set this translation vector to zero, as the transformation between hyper-surfaces can be fully characterized by the matrix $B$ (see Supplementary Material).


\begin{figure}[b]
\centering
\includegraphics[width=0.5\textwidth]{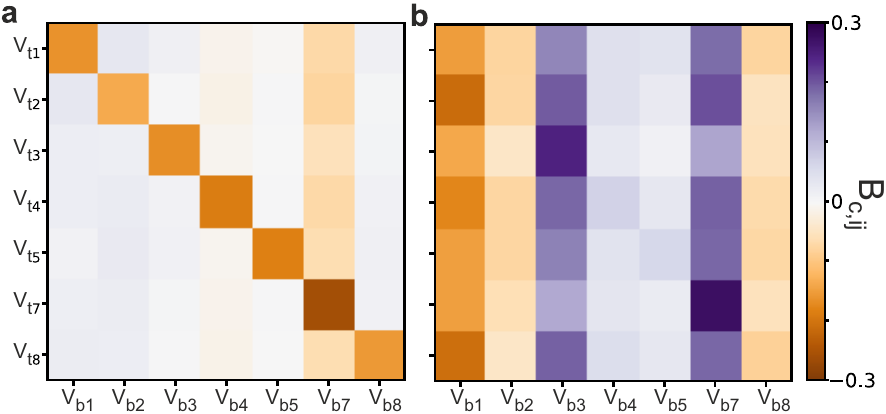}
\caption{\label{fig5}
\textbf{Learning about device variability.} $B_c$ matrices obtained using point set registration. Indices are the gate voltage coordinates $\mathbf{v_b}$ and $\mathbf{v_t}$. $V_6=0$~mV was fixed in Device 2 to prevent leakage currents. \textbf{a} Transformation between the hypersurface of Device 2 before and after a thermal cycle. \textbf{b} Transformation between the hyperfsurfaces of Device 1 and Device 2.}
\end{figure}

We have used this approach to quantify the variability between devices 1 and 2, and the effect of a thermal cycle in the hypersurface of Device 2. Figure~\ref{fig5} displays the matrix $B_c = B - I$ for each case, quantifying how much $B$, the transformation that converts a set of coordinates from one hypersurface onto the other, differs from the identity matrix (I). Non-zero elements of $B_c$ thus indicate device variability. Diagonal elements of $B_c$ are responsible for scale transformations and can be interpreted as a capacitance change for a given gate electrode. Off-diagonal elements are responsible for shear transformations and can be interpreted as a change in cross-capacitance between a pair of gate electrodes.



Fig.~\ref{fig5}a shows $B_c$ corresponding to the changes in the hypersurface of Device 2 after a thermal cycle. This transformation shows that device variability in a thermal cycle is dominated by a uniform change in capacitance for all gate electrodes. We have also measured $B_c$ for a thermal cycle of Device 1 (see Supplementary Material). Fig.~\ref{fig5}b displays $B_c$ comparing the hypersurface of Device 1 with the hypersurface of Device 2. We observe that the variability between these devices, which share a similar gate architecture, is given by non-uniform changes in gate electrode capacitance, as well as by changes in cross-capacitance. This variability is attributed to charge  traps  and  other device defects, such as a small differences in the patterning of gate electrodes.




\section{Conclusion}
We demonstrated an algorithm capable of tuning a quantum device with multiple gate electrodes in approximately 70~mins. This was achieved by efficiently navigating a multi dimensional parameter space without manual input or previous knowledge about the device architecture. This tuning time was reproduced in different runs of the algorithm, and in a different device with a similar gate architecture. 
Our tuning algorithm is able to tune devices with different number of gate electrodes with no modifications. We showed that gate electrodes with similar functions can be grouped to reduce the dimensionality of the gate voltage space and reduce tuning times to 36~mins. Tuning times might be further improved with efficient measurement techniques~\cite{Lennon2018}, as measurement and gate voltage ramping times were found to be the limiting factor. We analysed our algorithm design through an ablation study, which allowed us to justify and highlight the importance of each of its modules. The improvement over the pure random search without peak detection is estimated to be 179 times. 

We showed that device variability can be quantified using point set registration by uniform sampling of the hypersurface separating regions of high and low current in gate voltage space. We found that variability between devices with similar gate architectures is given by non-uniform changes in gate capacitances and cross-capacitances. Variability across thermal cycles is only given by a uniform change in gate capacitances. 


Other device architectures might use the sampling phase of our algorithm as a first tuning step, and the investigation phase can be adapted to tune quantum devices into more diverse configurations. To achieve full automated tuning of a singlet-triplet qubit, it will be necessary to go beyond this work by tuning the quantum dot tunnel barriers, identifying spin-selective transitions, and configuring the device for single-shot readout.




\appendix

\section{The score function as a classifier}
\label{classifier}

One of key strength of the proposed algorithm is that it does not require an ideal score function. It is important to highlight that we are using the score function just as a classifier, instead of aiming at finding the gate voltage configuration that maximises the score. The reason for this is threefold; (i) the score function is not always a smooth function; (ii) it does not always capture the quality of the transport features; (iii) it is just designed for a particular transport regime, in this case, honeycomb patterns. Therefore, the score threshold acts as a parameter that just controls the characteristics of the classifier. If the threshold is low, many high resolution scans not leading to double quantum dot transport features are produced. If the threshold is too high, then promising gate voltage windows are missed. The optimal threshold can be estimated by minimising the time required to produce a high-resolution current map that is labelled by humans as containing double quantum dot features.

\section{Mathematical analysis of ablation study results}
\label{ablation_appendix}

The results in the ablation study can be verified under a few assumptions by a mathematical derivation of $\mu_t$ (see Supplementary Material).  
From this derivation, we can compare the expected times $t^\textrm{abl}=t_{500}/500$ and $\mu_t^\textrm{abl}$ for \textit{Pure random}, \textit{Uniform surface}, and \textit{Peak selection}, 
\begin{align}
\label{eq:eq.1}
    t^\textrm{abl} &= t_\textrm{others}+P\left(\textrm{peaks}\right)t_\textrm{2D} \\
\label{eq:eq.2}
    \mu_t^\textrm{abl} &= \left(\frac{t_\textrm{others}}{P(\textrm{peaks})} + t_\textrm{2D} \right) \frac{1}{P(\textrm{success}|\textrm{peaks})},
\end{align}
where $t_\textrm{2D}=t_\textrm{2D-L}+t_\textrm{2D-H}$ with $t_\textrm{2D-L(H)}$ the expected measurement times corresponding to low (high) resolution current maps acquired in a sampling iteration. $t_\textrm{others}$ accounts for the rest of the investigation and sampling time, including ramping of gate voltages, peak detection, and computation time. The simulation of the Brownian particles is conducted in parallel with the investigation phase of the coordinate proposed by the sampler in a previous run, and it does not increase $t_\textrm{others}$. Note that the ablation study algorithms do not include the score function decision, so high resolution current maps are always acquired. In this case, low resolution maps are not useful, but we have still included $t_\textrm{2D-L}$ in $t_\textrm{2D}$ to keep the comparison between algorithms consistent.
 We estimate $t_\textrm{others}\sim35$~s, $t_\textrm{2D-L}\sim33$~s and $t_\textrm{2D-H}\sim273$~s.

Equations~(\ref{eq:eq.1}) and ~(\ref{eq:eq.2}) confirm our results in Fig.~\ref{fig4}a, as $t^\textrm{abl}$ grows for increasing $P(\textrm{peaks})$. They also show that the observed decrease in $\mu_t^\textrm{abl}$ is a consequence of an increase in $P(\textrm{peaks})$, given that $P(\textrm{success}|\textrm{peaks})$ remains mostly constant within experimental uncertainty (Fig.~\ref{fig4}b). $t_\textrm{2D}$ is constant and $t_\textrm{others}$ increases only slightly from \textit{Pure random} to \textit{Peak selection} (see Supplementary Material). We observe that optimising $t_{2D}$ is key to decrease $\mu_t$, motivating the introduction of a score function.

For the \textit{Full decision} algorithm, $\mu_t$ and $t$ are,
\begin{align*}
    t^\textrm{full}=& t_\textrm{others}+P\left(\textrm{peaks}\right)t_\textrm{2D-L} + P\left(\textrm{highres}\right)t_\textrm{2D-H} \\
    \mu_t^\textrm{full} =& \left(\frac{t_\textrm{others}}{P(\textrm{peaks})} + t_\textrm{2D-L}+ P(\textrm{highres}|\textrm{peaks})t_\textrm{2D-H}\right) \nonumber \\
    &\times\frac{1}{P(\textrm{success}|\textrm{peaks})},
\end{align*}
where $P(\textrm{highres})$ is the probability of acquiring a high resolution current map. The probability of acquiring a high resolution current map conditional on the observation of Coulomb peaks is $P(\textrm{highres}|\textrm{peaks})$.


The score function decision reduces $t$, because  $t^\textrm{abl}-t^\textrm{full}=P(\textrm{peaks})(1-P(\textrm{highres}|\textrm{peaks}))t_\textrm{2D-H}$ and $P(\textrm{highres}|\textrm{peaks})<1$. The reduction in $t$ given by the introduction of the score function decision is observed in Fig~\ref{fig4}a. 

Comparisons between $\mu_t^\textrm{abl}$ and $\mu_t^\textrm{full}$ can be affected by the dependence of $P(\textrm{success}|\textrm{peaks})$ on the score function threshold. In Fig.~\ref{fig4}b, however, we observe that $P(\textrm{success}|\textrm{peaks})$ is similar for $\textit{Peak selection}$ and $\textit{Full decision}$. This implies that the introduction of a score function threshold does not reduce the probability of success. 

In this case, 
\begin{equation*}
    \mu_t^\textrm{abl} - \mu_t^\textrm{full}=
    \frac{1-P(\textrm{highres}|\textrm{peaks})}{P(\textrm{success}|\textrm{peaks})}t_\textrm{2D-H}.
\end{equation*}
This equation confirms that that the score function reduces $\mu_t$ in the case that the score function threshold does not degrade $P(\textrm{success}|\textrm{peaks})$. Further analysis on the optimal threshold, i.e the threshold that minimizes $\mu_t^\textrm{full}$, can be found in Supplementary Material.


\subsection*{Acknowledgements}
We acknowledge J. Zimmerman and A. C. Gossard for the growth of the AlGaAs/GaAs heterostructure, F. Kuemmeth for useful discussions and Bobak Shahriari for proof reading the manuscript. This work was supported by the Royal Society, the EPSRC National Quantum Technology Hub in Networked Quantum Information Technology (EP/M013243/1), Quantum Technology Capital (EP/N014995/1), EPSRC Platform Grant (EP/R029229/1), the European Research Council (grant agreement 818751), the Swiss NSF Project 179024, the Swiss Nanoscience Institute, the NCCR QSIT and the EU H2020 European Microkelvin Platform EMP grant No. 824109. This publication was also made possible through support from Templeton World Charity Foundation and John Templeton Foundation. The opinions expressed in this publication are those of the authors and do not necessarily reflect the views of the Templeton Foundations. 

\subsection*{Author Contributions}
D.T.L., N.E., F.V., N.A. and the machine performed the experiments. H.M. developed the algorithm in collaboration with J.K., M.A.O and D.S. The sample was fabricated by L.C.C., L.Y., and D.M.Z. The project was conceived by G.A.D.B., E.A.L. and N.A. 


\subsection*{Competing Interests}
The authors declare that they have no
competing financial interests.
\subsection*{Correspondence}
Correspondence and requests for materials
should be addressed to Natalia Ares (email: natalia.ares@materials.ox.ac.uk).

%


\newcommand{\beginsupplement}{%
	\setcounter{table}{0}
	\renewcommand{\thetable}{S\arabic{table}}%
	\setcounter{figure}{0}
	\renewcommand{\thefigure}{S\arabic{figure}}%
	\renewcommand{\thesection}{*\Alph{section}}%
	\setcounter{equation}{0}
	\renewcommand{\theequation}{S\arabic{equation}}%
}

\beginsupplement

\clearpage
\section*{Supplementary Material}
\twocolumngrid
\subsection*{Hypersurface detector}

Finding the hypersurface boundary or `pinch-off' boundary requires an expected maximum and minimum current value. This is obtained by a single current trace before running the algorithm. The pinch-off is defined by a threshold, chosen as 20\% of the maximum current. The algorithm measures a trace in direction $\mathbf{u}$ from $\mathbf{o}$: $\mathbf{u}x_i + \mathbf{o}$, where $i$ is the step number and $x_i$ is the distance to the origin at the $i$th step. We choose $x_0=0$ and $x_{i+1}=x_i+10$~mV. For each step $i$, the pinch-off detector identifies the gate voltage coordinate for which the measured current is lower than the threshold. If the current continues to be below the threshold for a certain range $R$, set to 50mV for all experiments, the algorithm considers the device is pinched off.

\subsection*{Gaussian process models}
In this paper, we use two separate models that rely on Gaussian processes: one for estimating the hypersurface $r(\mathbf{u})$, and the other for estimating the location of Coulomb peaks $\tilde{P}_\mathrm{peak}(\mathbf{v})$. Each Gaussian process requires a prior distribution, which is defined by a prior mean function and a covariance kernel.
By setting the prior mean function $m_r(\cdot)$ and covariance kernel $k_r(\cdot, \cdot)$, the prior distribution is $\tilde{r}(\mathbf{u})\sim\mathcal{N}(m_r(\mathbf{u}), k_r(\mathbf{u},\mathbf{u}))$, and the covariance is $\mathrm{cov}[\tilde{r}(\mathbf{u}_1),\tilde{r}(\mathbf{u}_2)]=k_r(\mathbf{u}_1,\mathbf{u}_2)$. 

For the model of $r(\textbf{u})$ we initially have no prior knowledge, except for the maximum and minimum gate voltage bounds. Thus the range of $r$ is: $r\in [r_\mathrm{min}, r_\mathrm{max}]=[0, \sqrt{N}\times 2\mathrm{V}]$, where $N$ is the number of gate electrodes. We set the prior distribution in this range: $m_r(\mathbf{u})=(r_\mathrm{max}-r_\mathrm{min})/2$, $k_r(\mathbf{u},\mathbf{u})=(r_\mathrm{max}-r_\mathrm{min})^2/4^2$. This prior means that $r_\mathrm{max}$ and $r_\mathrm{min}$ will be two standard deviations away from $m_r(\mathbf{u})$. 

To model the probability $\tilde{P}_\mathrm{peak}(\mathbf{v})$, we factorise it into two components $\tilde{P}_\mathrm{peak}(\mathbf{v})= \tilde{P}_\mathrm{peak\vert valid}(\mathbf{v})\tilde{P}_\mathrm{valid}(\mathbf{v})$, where $\tilde{P}_\mathrm{valid}(\mathbf{v})$ is the probability that $\mathbf{v}$ corresponds to a pinch-off location, and not to the boundary of the operable gate voltages [0,-2]~V. Each component of $\tilde{P}_\mathrm{peak}(\mathbf{v})$ is estimated with a Gaussian process classification model.

For all Gaussian process models the Matern 5/2 kernel is used for the prior covariance. The prior covariance is defined by length scales $l_k$ with $k = 1,...,N$, setting the smoothness of the model predictions. These length scales are periodically optimised to the Maximum a posterior (MAP) using gathered data and a prior gamma distribution. The gamma distribution is set to have mean $0.4$, and variance $0.1^2$ for $r(\mathbf{u})$; mean $500$ and variance $100^2$ for $\tilde{P}_\mathrm{peak\vert valid}$; and mean $50$, and variance $20^2$ for $\tilde{P}_\mathrm{valid}$.

\subsection*{Score function}
The score function is designed to distinguish between measurements containing double quantum dot features, single quantum dot features, and measurements for which the confinement potential is not well defined. This could be achieved using modern machine learning techniques, however, the score function we use is not reliant on training data and hence is less device and labeller specific. The score consists of three separate scores multiplied: the orientation of gradients score, the sharpness score, and the fit direction score.


The orientation of gradient score aims to capture if Coulomb peaks appear as straight lines in a current map or if a honeycomb pattern is present. This is achieved by taking the numerical derivative of the current map and producing gradient unit vectors. These gradient vectors will exhibit a distribution of angles. If Coulomb peaks appear as straight lines, then the distribution of angles will reveal two high-density collections of unit vectors, separated by $\pi$ rad from each other. In this case, the high-density collections will have a small variance and a single line can fit the two points they define. For a honeycomb pattern, this variance will be larger or the high-density collections will not be well defined, and thus at least two lines will be required to fit the high-density collections. The value of the gradient orientation score is defined as the difference between the average of residuals corresponding to a fit through the high-density collections. Gradients smaller than a certain value are not considered for the fitting. This threshold was inferred from the current noise.


The sharpness score aims to prioritise `sharp' Coulomb peaks. To compute the sharpness score the current map is first spit up into 16 smaller tiles. Each title is segmented into regions of high current and low current using a simple threshold. Then, the second order derivative of the current as a function of gate voltage is computed. This second order derivative is averaged over the high current regions and its standard deviation is calculated. The score for each tile is defined as the product of these average and standard deviation values. The scores for the 16 tiles are then averaged together giving the final sharpness score.

The fit direction score aims to further differentiate between single and double quantum dot behavior. The current map is first spit up into 16 smaller tiles, as for the sharpness score. For each of these tiles, a lines is fit to the high-density collections of gradient vectors, in the same way as for the orientation of gradient score. From this fit, a gradient vector angle is assigned to the tile. For each row of tiles (in direction $\hat{V_e}$), the standard deviation of gradient vector angles is calculated. In this way, the array of $4\times4$ tiles is reduced to a $1\times4$ vector. The average of the elements in this vector, defines the fit direction score.

\subsection*{Labelling procedure}
Current maps that are measured by the algorithms in the tuning and ablation study sections, are labelled to determine if they contain features that indicate the device is tuned to the double quantum dot regime. This labelling is performed by human experts. To remove human bias, labelling was performed by 3 independent labellers. 



Two separate data sets were labelled. The first contained 2048 current maps taken from all experiments presented in this paper, except the pure random experiment in section III A. The second contained 151 current maps, 51 of which were taken from the pure random experiment in section III A and 100 were randomly selected from the first data set. The current maps in each data set were randomly shuffled so that labellers could not identify which experiment produced a given current map. The labellers were the same for both data sets.

\subsection*{Point set registration}

Coherent point drift~\cite{Myronenko2010} was our choice of point set registration, since it can be used for point sets of high dimensionality. It is also simple to implement, it is robust to noise in the point sets, and supports affine transformations as well as rigid and non-rigid transformations. We focus on affine transformations  $T(x) = Bx + t$, where $B$ is a square matrix, and $t$ is a column vector. Reference ~\cite{Myronenko2010} presents the mathematical derivation for this case. In our application of the affine transformation, $t$ does not have significance in terms of the device physics, and it was therefore set to 0. A derivation for the case $t = 0$ was required. This was achieved by setting the centered point matrices $\hat{X} = X$ and $\hat{Y} = Y$ in the derivation from section 4 (Rigid \& affine point set registration) in Ref.~\cite{Myronenko2010}.


\subsection*{Bayesian statistics}
\label{subsec:Bayesian}

\subsubsection*{Single-labeller statistics}

To infer the expected tuning time $\mu_t$, and the probabilities $P(\textrm{peaks})$, and $P(\textrm{success}|\textrm{peaks})$, we use Bayesian inference with Jeffreys prior. 
Let $p$ denote the probability (either $P(\textrm{peaks})$ or $P(\textrm{success}|\textrm{peaks})$) that we want to estimate. The Jeffreys prior for a binomial distribution is $p\sim\textrm{Beta}(0.5,0.5)$. If we observe $k$ successes over $n$ trials, then the posterior distribution of $p$ is $p|k,n\sim\textrm{Beta}(0.5+k,0.5+n-k)$.

To infer $\mu_t$, we assume that the rate of success over time follows a Poisson distribution $k\sim\textrm{Poisson}(\lambda t_\mathrm{tot})$, where $t_\mathrm{tot}$ is the total time for an algorithm run, and $\lambda$ is a rate parameter. The expected time between two consecutive successes (i.e. the acquisition of currents maps displaying double quantum dot transport features) is $\mu_t=1/\lambda$. The Jeffreys prior for a Poisson distribution is $\lambda\sim\textrm{Gamma}(0.5,0)$, hence the posterior is $\lambda|k,t_\mathrm{tot}\sim\textrm{Gamma}(0.5+k,t_\mathrm{tot})$. Consequently, $\mu_t|k,t_\mathrm{tot}\sim\textrm{Inv-Gamma}(0.5+k,t_\mathrm{tot})$.

\subsubsection*{Multi-labeller statistics}

Before introducing multi-labeller statistics, let us introduce necessary notation: $\mathcal{D}=\lbrace(\mathbf{v}_i,Y_i)|i=1\sim n\rbrace$, where $\mathbf{v}_i$ is a voltage configuration, and $Y_i$ is a high-resolution scan in the vicinity of $\mathbf{v}_i$. If no high-resolution scan was acquired at $\mathbf{v}_i$, then $Y_i$ is an empty array. The data $\mathcal{D}$ requires a labelling function $\psi$: $\psi(Y_i)=0$ if there is no recognized double quantum dot transport features, or $1$ otherwise. But labellers might not agree, and thus we need to marginalise this effect. Let $\theta$ denote a parameter to be inferred ($\mu_t$, $\mathrm{P(peaks)}$, or $\mathrm{P(success|peaks)}$), then $p(\theta|\mathcal{D},\psi)=p(\mathcal{D}|\theta,\psi)p(\theta)/p(\mathcal{D})$, where $\Psi$ is a set of labellers. Note that $\psi$ only affects the likelihood. To minimize the affect of $\psi$, the posterior can be marginalised over $\psi$ and also approximated by samples of $\psi$:
\begin{equation*}
p(\theta|\mathcal{D})=\mathbb{E}_\psi [p(\theta|\mathcal{D},\psi)]\approx \frac{1}{|\Psi|} \sum_{\psi \in \Psi} p(\theta|\mathcal{D},\psi),
\end{equation*}
where $|\Psi|$ is the number of labellers. Therefore, the cumulative distribution function of $\theta|\mathcal{D}$ is
\begin{equation*}
P(\theta<z|\mathcal{D}) \approx \frac{1}{|\Psi|} \sum_{\psi\in\Psi} P(\theta<z|\mathcal{D},\psi).
\end{equation*}

\subsection*{Mathematical derivation of $\mu_t$}
We define $t_i$ as the time taken to produce the $i$th sample, and $\bar{t}_i=\sum_{j=1}^{i}t_j$. We also define $t_s$ as the interval between successes, or from start of the algorithm run to the first success if there were no previous success. Likewise, $n_s$ is the iteration number corresponding to the first success. We want to derive $E[t_s]$. For brevity, we define shortened notations: $\alpha=P(\textrm{peaks})$, $\beta=P(\textrm{success}|\textrm{peaks})$. Note that $P(\textrm{success})=\alpha\beta$.

Then the distribution of $t_s$ is
\begin{equation*}
P(t_s<t)=\sum_{i=1}^n  P(n_s=i)
P(\bar{t}_i<t|n_s=n).
\end{equation*}
The expected time is
\begin{align}
E[t_s]
&= \sum_{i=1}^n P(n_s=i)  E[\bar{t}_i|n_s=i] \nonumber \\
&= \sum_{i=1}^n (1-\alpha\beta)^{i-1}\alpha\beta E[\bar{t}_i|n_s=i].
\label{eqn:expectedtime}
\end{align}
The last term $E[\bar{t}_i|n_s=i]$ depends on the tuning algorithm.

\subsubsection*{Derivation of $\mu_t^\textrm{abl}$}

The time for each iteration of the algorithm is $t_n=t_{n,\textrm{other}} + t_{n,\textrm{2D}}\mathbb{I}_{n,\textrm{peaks}}$, where $\mathbb{I}_{n,\textrm{peaks}}$ is equal to 1 if peaks are detected at $\mathbf{v}(\mathbf{u}_n)$ and 0 otherwise, $t_{n,\textrm{2D}}$ and $t_{n,\textrm{other}}$ is the time taken for 2D scans and other related times for iteration $n$. Consequently, $\bar{t}_n=\sum_{i=1}^n t_{i,\textrm{other}} + \sum_{i=1}^n t_{i,\textrm{2D}}\mathbb{I}_{i,\textrm{peaks}}$, and $\bar{t}_n|(n_s=n) = \sum_{i=1}^n t_{i,\textrm{other}} + \sum_{i=1}^{n-1} t_{i,\textrm{2D}}\mathbb{I}_{i,\textrm{peaks}} + t_{n,\textrm{2D}}$, because $n_s=n$ implies that 2D scans are measured at $n$. We assume that $t_{i,\textrm{other}}$ and $t_{i,\textrm{2D}}$ are i.i.d. across $i$, and we denote the expected time $t_\textrm{others}$ and $t_\textrm{2D}$, respectively. Then the conditional expectation is
\begin{align}
E[\bar{t}_n|n_s=n] 
&=nt_\textrm{others} + t_\textrm{2D} + (n-1)P(\textrm{peaks}|\textrm{fail at }n)t_\textrm{2D}\nonumber\\
&=nt_\textrm{others} + t_\textrm{2D} + (n-1)\frac{(1-\beta)\alpha}{1-\alpha\beta}t_\textrm{2D},
\label{eqn:prob_conditional}
\end{align}
because $P(\textrm{peaks}|\textrm{fail at }n)=\frac{P(\textrm{fail at } n|\textrm{peaks})P(\textrm{peaks})}{P(\textrm{fail})}=\frac{(1-\beta)\alpha}{1-\alpha\beta}$. Substituting (\ref{eqn:prob_conditional}) into (\ref{eqn:expectedtime}) results in
\begin{equation*}
E[t_s] = \frac{t_\textrm{others}}{\alpha\beta} + \frac{t_\textrm{2D}}{\beta},
\end{equation*}
because of the following:
\begin{align*}
&\alpha\beta\sum_{n=1}^\infty (1-\alpha\beta)^{n-1} =1 \\
&\alpha\beta\sum_{n=1}^\infty (1-\alpha\beta)^{n-1}n =\frac{1}{\alpha\beta} \\
&\alpha\beta\sum_{n=1}^\infty (1-\alpha\beta)^{n-1}(n-1) =\frac{1-\alpha\beta}{\alpha\beta}.
\end{align*}

\subsubsection*{Derivation of $\mu_t^\textrm{full}$}

For brevity, we use the following shortened notations: $\alpha'=P(\textrm{peaks},\textrm{highres})=P(\textrm{highres})$, $\beta'=P(\textrm{success}|\textrm{peaks},\textrm{highres})=P(\textrm{success}|\textrm{highres})$. Deriving the conditional expectation is very similar to the previous section,

\begin{align*}
E[\bar{t}_n|n_s=n] 
=& nt_\textrm{others} + t_\textrm{2D-L} + (n-1)\frac{(1-\beta)\alpha}{1-\alpha\beta}t_\textrm{2D-L} + \\
&t_\textrm{2D-H} + (n-1)\frac{(1-\beta')\alpha'}{1-\alpha'\beta'}t_\textrm{2D-H}.
\label{eqn:prob_conditional_full}
\end{align*} Therefore, the marginalised expectation is
\begin{equation}
E[t_s] = \frac{t_\textrm{others}}{\alpha\beta} + \frac{t_\textrm{2D-L}}{\beta} + \frac{t_\textrm{2D-H}}{\beta'}.
\label{eqn:exptected_full}
\end{equation}
Note that $P(\textrm{success})=\alpha\beta=\alpha'\beta'$, which leads to $\beta'=\alpha\beta/\alpha'$. Also, $\alpha/\alpha'= P(\textrm{peaks})/P(\textrm{highres})=1/P(\textrm{highres}|\textrm{peaks})$. Substituting these yields
\begin{equation*}
E[t_s] = \frac{t_\textrm{others}}{\alpha\beta} + \frac{t_\textrm{2D-L}}{\beta} + \frac{P(\textrm{highres}|\textrm{peaks})t_\textrm{2D-H}}{\beta}.
\end{equation*}

\subsection*{Optimal threshold $\alpha'$}

An alternative expression for (\ref{eqn:exptected_full}) is
\begin{equation}
E[t_s|\alpha',\beta'] = \frac{1}{\beta'} \left( \frac{t_\textrm{OL}}{\alpha'}  + t_\textrm{2D-H} \right),
\label{eqn:expectation_q}
\end{equation}
where $t_\textrm{OL}=t_\textrm{others}+\alpha t_\textrm{2D-L}$. Note that $\beta'$ is the recall, which is the proportion of the real double dot samples over all positively classified samples, of a classifier, and it is related with $\alpha'$. Therefore, a model of $\beta'$ in terms of $\alpha'$ is required.

\subsubsection*{Perfect score function}
We define a score function is perfect if there exists a threshold that perfectly classifies double-dot samples from others. In this case, The model for $\beta'$ is
$$
\beta' = \min (1, q/\alpha') ,
$$
where $q$ is the actual proportion of double-dot samples. In other words, the recall is 1 if $\alpha'\leq q$, or $q/\alpha'$ otherwise. It is trivial to see that the expected time (\ref{eqn:expectation_q}) is minimum at $\alpha'=q$. If $q$ is unknown, which is usual, then we can utilise a prior distribution on $q$ and optimise $\alpha'$ given the prior distribution.

In summary, from this analysis we conclude that it is beneficial to keep the threshold as big as possible if the recall does not drop significantly. Also, the incentive of using a score function as classifier is higher when the relative cost of a high resolution scan is big.

\subsubsection*{Imperfect score function}
Let us assume that a score function is perfect with the probability $\xi$, or the score is random with the probability $(1-\xi)$. Then the model for $\beta'$ is
$$
\beta' = \frac{\min(q,\alpha')\xi + \alpha'q(1-\xi)}{\alpha'} .
$$
Again, it is trivial that the expected time (\ref{eqn:expectation_q}) increases when $\alpha'$ increases if $\alpha' \leq q$. 

In case of $\alpha' > q$, (\ref{eqn:expectation_q}) becomes
$$
E[t_s|\alpha'] = \frac{1}{q} \frac{t_\textrm{OL}+\alpha't_\textrm{2D-H}}{\alpha'(1-\xi)+\xi}.
$$
and the partial derivative w.r.t. $\alpha'$ is
$$
\frac{\partial E[t_s|\alpha']}{\partial \alpha'} =
\frac{1}{q} \frac{t_\textrm{2D-H}\xi - (1-\xi)t_\textrm{OL}}{\lbrace \alpha'(1-\xi)+\xi\rbrace^2}.
$$
Note that $\alpha'$ does not change the sign of the derivative, but $\xi$ changes the sign. Rearranging $\frac{\partial E[t_s|\alpha']}{\partial \alpha'}=0$ yields $\xi=\frac{t_\textrm{OL}}{t_\textrm{OL}+t_\textrm{2D-H}}$. It means that the optimal $\alpha'$ is $q$ if $\xi>\frac{t_\textrm{OL}}{t_\textrm{OL}+t_\textrm{2D-H}}$. Otherwise, increasing $\alpha'$ always decreases the expected time, hence setting $\alpha'$ as big as possible reduces the time. It means that there is no advantage of using the score function if $\xi<\frac{t_\textrm{OL}}{t_\textrm{OL}+t_\textrm{2D-H}}$.

\subsection*{Additional results and raw data}

\begin{figure*}
	\centering
	\includegraphics[width=0.8\textwidth]{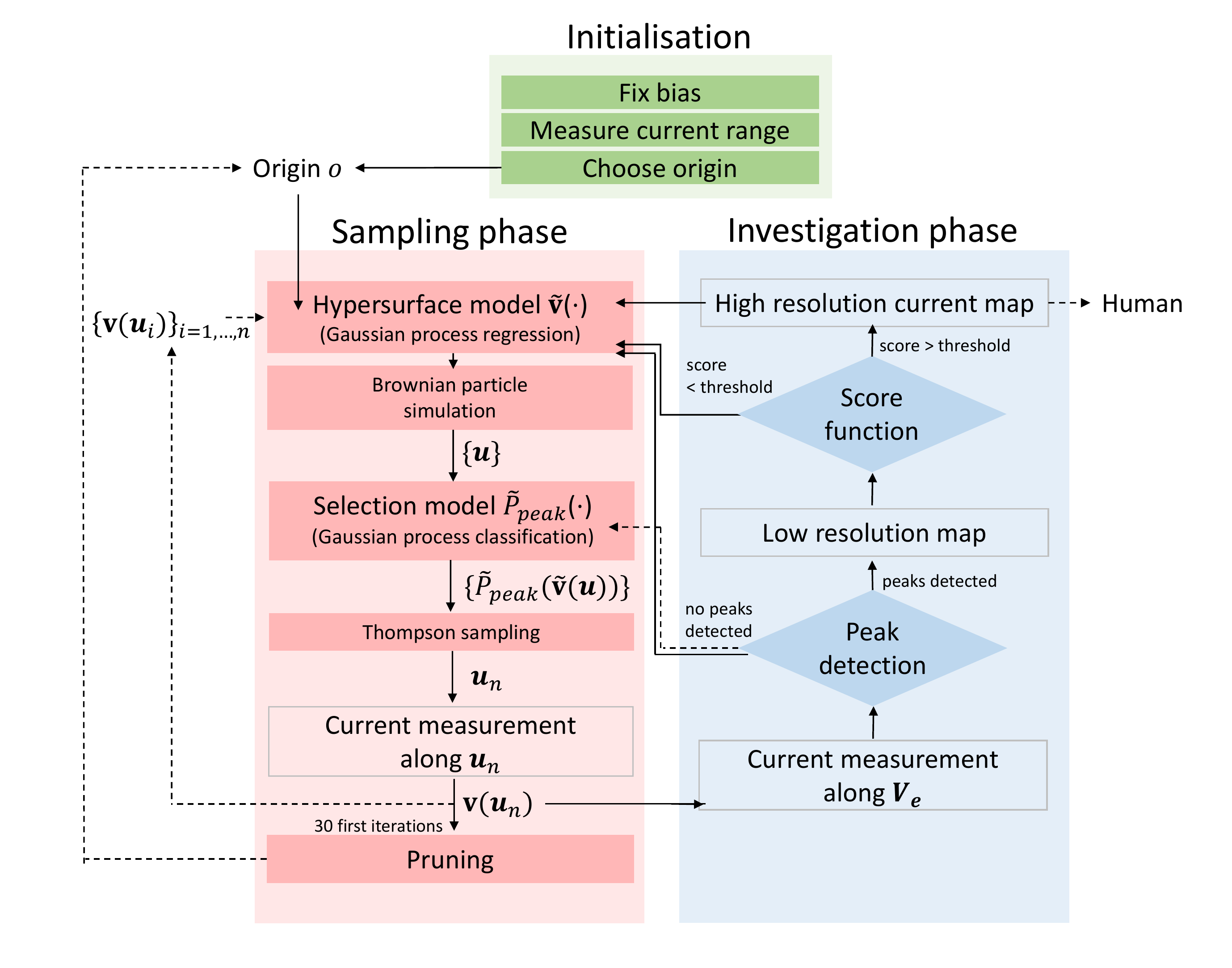}[b]
	\caption{\label{sup1}
		\textbf{Detailed schematic of the entire algorithm.} Solid lines are control flows, and dotted lines are information flows. Variables are defined in the main text.}
\end{figure*}

\begin{figure}
	\centering
	\includegraphics[width=0.4\textwidth]{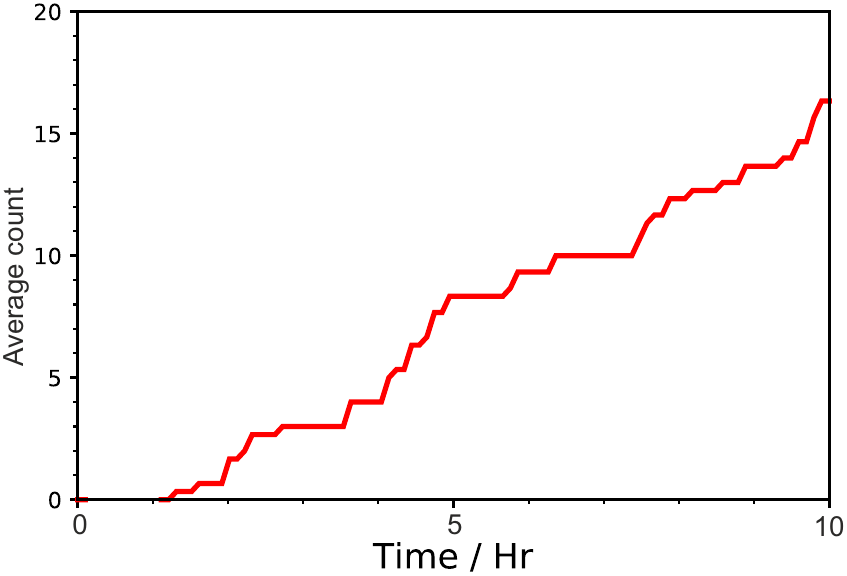}
	\caption{\label{sup2}
		\textbf{Tuning performance of the grouped gate electrode variant of the algorithm.} Average count of current maps displaying double quantum dot features, as a function of laboratory time for the grouped gates implementation of the algorithm. Current maps are labelled by humans a posteriori, i.e. after the algorithm is stopped.}
\end{figure}

\begin{figure}
	\centering
	\includegraphics[width=0.8\linewidth]{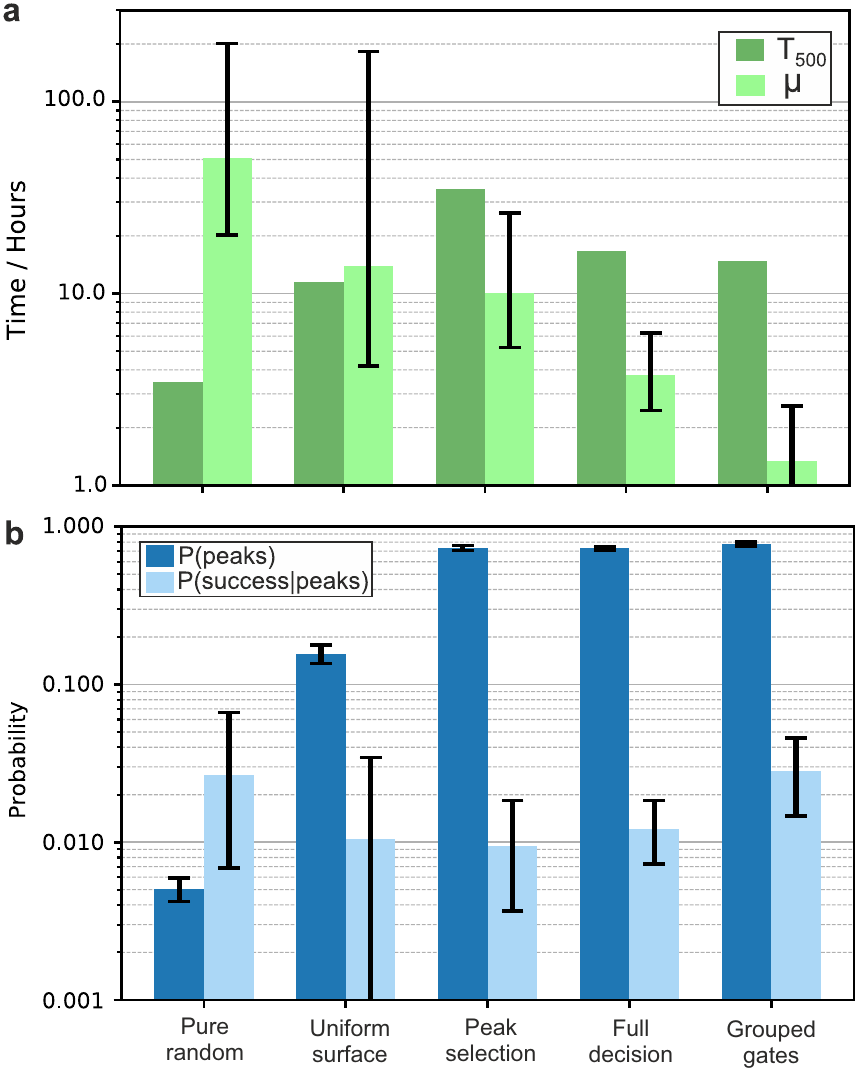}
	\caption{\label{sup4}
		\textbf{Additional ablation study data.} \textbf{a} and \textbf{b} bar charts comparing $\mu_t$ (light green), $t_{\text{500}}$ (dark green), $P(\textrm{peaks})$ (dark blue) and $P(\textrm{success}|\textrm{peaks})$ (light blue) for the different algorithms considered in the ablation study with the inclusion of the grouped gates implementation. Error bars represent $10\%$ and $90\%$ confidence intervals.}
\end{figure}

\begin{figure}
	\centering
	\includegraphics[width=0.3\textwidth]{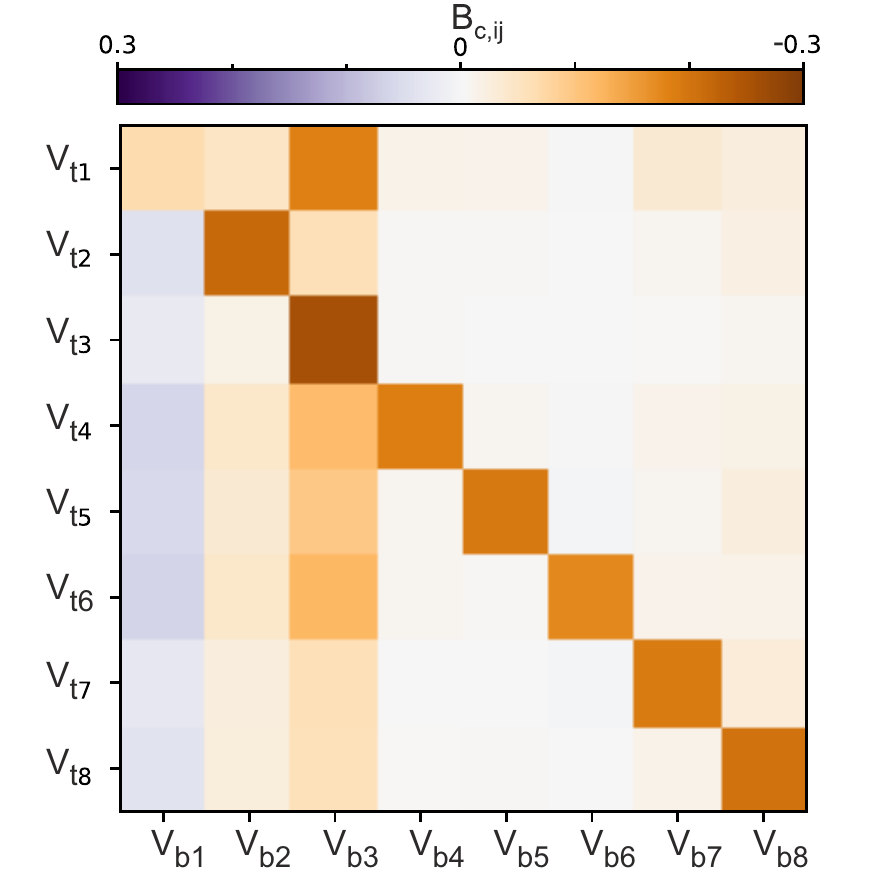}
	\caption{\label{sup3}
		\textbf{Additional device variability data.} The plot shows $B_c = B - I$ learned via point set registration from point sets $v_a$ and $v_b$ acquired on Device 1 before and after a thermal cycle.}
\end{figure}

Figure~\ref{sup1} is a detailed block diagram of the algorithm. Figure~\ref{sup2} shows the tuning performance, on device 1, of the grouped gates implementation of the algorithm, under the same conditions as Section IIIA in the main text. We have also included the grouped gates implementation of the algorithm, under the same conditions as in Section IIIB, in Fig.~\ref{sup4}. Additional results for the device variability study in Section IIIC, where device 1 is thermal cycled, is displayed in Fig.~\ref{sup3}. The raw data used to calculate tuning performance in Section IIIA is shown in Table S1. The raw data used to calculate tuning performance in Section IIIB is shown in Table S2.

\onecolumngrid
\FloatBarrier
\begin{table}
	\begin{center}
		\begin{tabular}{||c c c c c c||} 
			\hline
			Experiment & Iterations  & Time (hours) & Labeller 1 & Labeller 2 & Labeller 3 \\ [0.5ex] 
			\hline
			Device1, run1 & 500 & 8.2 & 7 & 8 & 7 \\ 
			Device1, run2 & 500 & 9.1 & 7 & 8 & 6 \\ 
			Device1, run3 & 500 & 8.2 & 5 & 4 & 6 \\ 
			Device1, run4 & 500 & 8.4 & 5 & 11 & 10 \\ 
			Device1, run5 & 500 & 7.7 & 4 & 6 & 4 \\ 
			\hline
			Device2, run1 & 500 & 11.2 & 2 & 3 & 1 \\ 
			Device2, run2 & 500 & 11.4 & 1 & 6 & 6 \\ 
			Device2, run3 & 500 & 11.2 & 0 & 4 & 2 \\ 
			Device2, run4 & 500 & 11.4 & 3 & 9 & 7 \\ 
			Device2, run5 & 500 & 10.5 & 3 & 4 & 5 \\ 
			\hline
			Device2, Grouped gates, run1 & 500 & 11.1 & 12 & 19 & 20 \\ 
			\hline
		\end{tabular}
		\caption{Labelling result of the experiments in section~\ref{subsec:result_tuning}. It shows the number of samples classified as successes for each labeller.}
		\label{table:tuning}
	\end{center}
\end{table}

\begin{table}
	\begin{center}
		\begin{tabular}{||c c c c c c||} 
			\hline
			Experiment & Iterations  & Time (hours) & Labeller 1 & Labeller 2 & Labeller 3 \\ [0.5ex] 
			\hline
			Pure random, run1 & 1000 & 6.9 & 0 & 0 & 0 \\ 
			Pure random, run2 & 10000 & 68.9 & 1 & 2 & 1 \\ 
			\hline
			Uniform surface, run1 & 500 & 11.5 & 0 & 1 & 1 \\ 
			\hline
			High-res always, run1 & 500 & 35.3 & 4 & 4 & 2 \\ 
			\hline
			Full method, run1 & 500 & 15.7 & 2 & 6 & 5 \\ 
			Full method, run2 & 459 & 16.2 & 5 & 3 & 4 \\ 
			\hline
			Full method + grouped gates, run1 & 500 & 14.8 & 7 & 15 & 11 \\ 
			\hline
		\end{tabular}
		\caption{Labelling result of the experiments in section~\ref{subsec:result_abl}. It shows the number of samples classified as successes for each labeller. The second run of \textit{Full method} could not reach 500 iterations, because the algorithm was interrupted by some external noise.}
		\label{table:abl}
	\end{center}
\end{table}


\begin{thebibliography}{22}
\bibitem [{\citenamefont {Kuemmeth}()}]{humantuning}%
	\BibitemOpen
	\bibfield  {author} {\bibinfo {author} {\bibfnamefont {F.}~\bibnamefont
			{Kuemmeth}},\ }\href@noop {} {}\bibinfo {howpublished} {personal
		communication. Others whom we have consulted cite similar values}\BibitemShut {NoStop}%
\end{thebibliography}
\end{document}